\documentclass[a4paper,11pt]{article}
\makeatletter
\providecommand\input@path{}
\g@addto@macro\input@path{{tex/}}
\makeatother
\usepackage{jheppub}

\usepackage[T1]{fontenc}
\usepackage[utf8]{inputenc}
\usepackage{textcomp}
\ifdefined\DeclareUnicodeCharacter
\DeclareUnicodeCharacter{03C4}{\ensuremath{\tau}}
\DeclareUnicodeCharacter{2212}{\ensuremath{-}}
\fi
\usepackage{graphicx}
\usepackage{xcolor}
\usepackage{placeins}
\hypersetup{
  pdftitle={Four-loop electron EDM matching and timelike tau electric dipole form factor},
  pdfauthor={Chia-Wei Liu and Zi-Yue Zou},
  pdfsubject={Source document date: 15 October 2026}
}
\allowdisplaybreaks

\title{Four-loop electron EDM matching and timelike \texorpdfstring{$\tau$}{tau} electric dipole form factor}
\author[a]{\href{https://orcid.org/0000-0002-3969-6119}{Chia-Wei Liu}}
\emailAdd{chiaweiliu@ucas.ac.cn}
\affiliation[a]{School of Fundamental Physics and Mathematical Sciences, Hangzhou Institute for Advanced Study, UCAS, Hangzhou 310024, China}
\author[b,c]{\href{https://orcid.org/0009-0004-4446-527X}{Zi-Yue Zou}}
\emailAdd{ziy\_zou@sjtu.edu.cn}
\affiliation[b]{State Key Laboratory of Dark Matter Physics, Tsung-Dao Lee Institute \& School of Physics and Astronomy, Shanghai Jiao Tong University, Shanghai 200240, China}
\affiliation[c]{Key Laboratory for Particle Astrophysics and Cosmology (MOE) \& Shanghai Key Laboratory for Particle Physics and Cosmology, Tsung-Dao Lee Institute \& School of Physics and Astronomy, Shanghai Jiao Tong University, Shanghai 200240, China}

\abstract{
We compute the four-loop matching of a CP-odd four-$\tau$ interaction onto the electron electric dipole moment (EDM), together with the leading timelike $\tau$ dipole form factor generated by the same source.
Both results are scale independent at this order.
Without a local electron dipole, exact electron-EDM cancellation still leaves a timelike $\tau$ signal proportional to the four-$\tau$ coefficient.
Its absorptive part cannot be canceled by a real local dipole, allowing collider limits to constrain the cancellation direction and complement static electron-EDM searches.
}

\begin{document}
\maketitle

\section{Introduction}

Electric dipole moments (EDMs) probe CP violation in lepton interactions.
Precision electron-EDM searches constrain heavy-lepton sources through radiative effects~\cite{Chang:2005wu,Ema:2022wxd}, while $e^+e^-\to\tau^+\tau^-$ provides direct access to the $\tau$ electric dipole form factor.
Spin and decay correlations probe its real and imaginary parts~\cite{Bernreuther:2021elu,He:2025ewk}, including absorptive signals from GeV-scale new physics~\cite{Huang:2025ghw}.
A consistent interpretation requires control of radiative corrections~\cite{Gogniat:2025eom}.

Among the charged leptons, the $\tau$ EDM is the least constrained, and its chirality-flip enhancement makes it a sensitive probe of CP violation in the third generation~\cite{Huang:2025ghw}.
A GeV-scale CP-violating spin-zero mediator with both scalar and pseudoscalar couplings to the $\tau$ can induce sizable real and imaginary parts of the timelike $\tau$ EDM form factor~\cite{Huang:2025ghw,Hoferichter:2025zjp}.
Such signals escape the indirect bound from the electron EDM, which assumes a real, momentum-independent $\tau$ form factor~\cite{Ema:2022wxd,Huang:2025ghw}.
When the mediator is heavier than the collision energy but lighter than the electroweak scale, it can be integrated out, and its effects are described model independently by the low-energy effective field theory (EFT) below the electroweak scale~\cite{Jenkins:2017jig,Jenkins:2017dyc,Panico:2018hal}.
The resulting CP-odd interactions are
\begin{equation}
\begin{aligned}
\mathcal{L}_{\mathrm{int}}&=C_{SP}(\mu)(\overline{\tau}\tau)(\overline{\tau}i\gamma^5\tau) -\frac{i}{2}d_\tau(\mu)\overline{\tau}\sigma^{\mu\nu}\gamma^5\tau F_{\mu\nu} {-\frac{i}{2}d_e(\mu)\overline{e}\sigma^{\mu\nu}\gamma^5e F_{\mu\nu}}.
\end{aligned}
\label{eq:cp-odd-interactions}
\end{equation}
with $\sigma^{\mu\nu}=i[\gamma^\mu,\gamma^\nu]/2$.
Hermiticity requires real $C_{SP}(\mu)$ and $d_\tau(\mu)$, and the scalar--pseudoscalar four-$\tau$ operator is CP odd.
{The electron-dipole counterterm absorbs the remaining $C_{SP}$-induced UV divergence.
The real renormalized coefficient $d_e(\mu)$ is an independent parameter; we impose $d_e(m_\tau)=0$ in the scheme specified below.}
Tree-level exchange of the mediator generates the four-$\tau$ operator, whereas the $\tau$ dipole first arises at one loop~\cite{Lebedev:2002ne,DiLuzio:2020oah}.
Four-lepton operators are therefore the leading low-energy imprint of such CP-violating scalar sectors, and the EFT relates their timelike $\tau$ signal and the static electron EDM within a single framework.

Interpreting timelike measurements in terms of $\tau$ dipole moments also requires the non-dipole operators of this EFT~\cite{Aebischer:2021uvt,Hu:2024dnb}.
The measured quantity is the dipole form factor at the collision energy, into which four-fermion operators enter at one loop~\cite{Aebischer:2021uvt}; above the corresponding pair threshold, they generate absorptive parts that spin-asymmetry measurements can detect~\cite{Gogniat:2026zvf}.
For the magnetic moment, this route could even provide the strongest constraints on four-$\tau$ operators~\cite{Gogniat:2026zvf}.
In the CP-odd sector considered here, a real local $\tau$ dipole contributes at tree level, whereas a CP-odd scalar--pseudoscalar four-$\tau$ interaction generates a $\tau$-pair cut at one loop, whose momentum dependence and absorptive part a real local dipole cannot reproduce~\cite{Du:2026yoi}.

If the local $\tau$ dipole is counted at the order of the one-loop four-$\tau$ contribution, its three-loop transfer to the electron EDM~\cite{Ema:2022wxd} and the four-loop four-$\tau$ matching must both be retained.
We compute the latter at leading power in the electron-to-$\tau$ mass ratio and combine it with the leading timelike $\tau$ form factor.

\section{\texorpdfstring{$\tau$}{tau} dipole form factor}

We work in low-energy quantum electrodynamics (QED), with pole masses $m_e$ and $m_\tau$ and the on-shell (OS) charge $\mathrm{e}>0$.
We adopt the radiative counting $d_\tau=\mathcal{O}(\mathrm{e}m_\tau C_{SP}/(8\pi^2))$.
The local dipole and four-$\tau$ loop then contribute at the same source order, while their finite matching boundary values remain independent; operator mixing fixes the running, not their relative normalization~\cite{Jenkins:2017jig,Jenkins:2017dyc}.
For new physics above the electroweak scale, the scalar four-$\tau$ contact first appears at dimension eight, whereas the dipole is dimension six~\cite{Panico:2018hal,Aebischer:2021uvt,Du:2026yoi}, so this counting applies to mediators below the electroweak scale.

\begin{figure}[tbp]
\centering
\includegraphics[width=\linewidth]{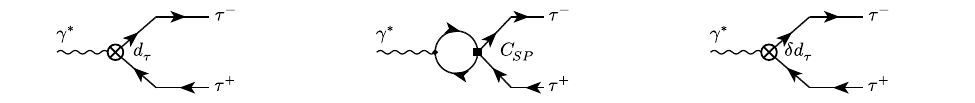}\\[2pt]
\makebox[\linewidth][l]{\hspace*{0.147\linewidth}\makebox[0pt]{(a)}\hspace*{0.328\linewidth}\makebox[0pt]{(b)}\hspace*{0.343\linewidth}\makebox[0pt]{(c)}}
\caption{Leading $\tau$ current.
Panel (a): local dipole; panel (b): four-$\tau$ exchange bubble; panel (c): mixing counterterm.
Squares denote $C_{SP}$, dots QED vertices, and crossed circles dipoles.
Direct photon contractions vanish; the pole mass is real.}
\label{fig:tau-leading}
\end{figure}

At fixed OS charge and pole mass, the one-loop equations, linear in the Wilson coefficients, are
\begin{equation}
\frac{\mathrm{d}d_\tau(\mu)}{\mathrm{d}\log \mu } =\frac{5\alpha}{2\pi}d_\tau(\mu)+\frac{\mathrm{e}m_\tau}{4\pi^2}C_{SP}(\mu), ~~~ \frac{\mathrm{d}C_{SP}(\mu)}{\mathrm{d}\log \mu } =\frac{3\alpha}{\pi}C_{SP}(\mu),
\label{dipole-rg}
\end{equation}
where $\alpha=\mathrm{e}^2/(4\pi)$.
The current in Fig.~\ref{fig:tau-leading} fixes the source-mixing term in the first relation in Eq.~\eqref{dipole-rg}; its homogeneous term arises from QED dressing of the local dipole.
The second relation uses the four-$\tau$ anomalous dimension of Ref.~\cite{Jenkins:2017dyc}, Eq.~(C.55).
We drop the $\alpha$ terms in Eq.~\eqref{dipole-rg} and two-loop source terms, since they enter one source order higher.
Choosing the reference scale $\mu_0=m_\tau$ gives
\begin{equation}
d_\tau(\mu)=d_\tau(m_\tau) +\frac{\mathrm{e}m_\tau C_{SP}(m_\tau)}{4\pi^2}\log\frac{\mu}{m_\tau},
\label{leading-evolution}
\end{equation}
with $C_{SP}(\mu)=C_{SP}(m_\tau)$ at this order.

For $\gamma^*(q)\to\tau^-(p_-)+\tau^+(p_+)$, the photon momentum $q$ is incoming and both $\tau$ momenta are outgoing, so $q=p_-+p_+$, $s=q^2$, and $p_-^2=p_+^2=m_\tau^2$.
With on-shell $\tau$ legs, the electromagnetic vertex can be decomposed into form factors~\cite{Bernreuther:2021elu,Du:2026yoi}.
We define the amputated vertex Feynman rule as $-i\Gamma^\mu$, with $Q_\tau$ the $\tau$ charge in units of $\mathrm{e}$.
Keeping the tree-level vector term and defining the electric dipole form factor $F_3(s)$, we write
\begin{equation}
\Gamma^\mu(s)=\mathrm{e}Q_\tau\gamma^\mu+\mathrm{e}F_3(s)\sigma^{\mu\nu}q_\nu\gamma^5.
\label{formfactor}
\end{equation}
The $\tau$ current is $\overline{u}(p_-)\Gamma^\mu(s)v(p_+)$.
The EDM form factor is $-\mathrm{e}F_3(s)$; its value at $s=0$ is the physical static EDM.
We are interested only in $\tau$-pair production above threshold, $s>4m_\tau^2$, and define $\beta=\sqrt{1-4m_\tau^2/s}$.
Combining the local dipole and the renormalized four-$\tau$ loop in Fig.~\ref{fig:tau-leading}, and using Eq.~\eqref{leading-evolution} to express the coefficients at $\mu=m_\tau$, gives
\begin{equation}
F_3(s)=-\frac{d_\tau(m_\tau)}{\mathrm{e}} +\frac{m_\tau C_{SP}(m_\tau)}{8\pi^2}\left[2-\beta\log\frac{1+\beta}{1-\beta}+i\pi\beta\right].
\label{tau-leading}
\end{equation}
The $i0$ prescription fixes the sign of the absorptive part.
The $\tau$-pair cut gives an absorptive part despite real local coefficients; this nonanalytic momentum dependence cannot be absorbed into the local dipole.

\section{Electron EDM matching and cancellation}

The three-loop transfer of a local heavy-lepton EDM to the electron EDM was studied in Refs.~\cite{Grozin:2008nw,Grozin:2009jq}.
The electron external-momentum terms at the same leading power were subsequently included in Ref.~\cite{Ema:2022wxd}, giving the coefficient $19\zeta(3)/4-71/24$.
Our calculation reproduces this coefficient.

We use dimensional regularization with $D=4-2\epsilon$ and the Breitenlohner--Maison--'t~Hooft--Veltman (BMHV) prescription for $\gamma^5$~\cite{tHooft:1972tcz,Breitenlohner:1977hr}: $g=\bar g+\hat g$, with barred indices spanning four dimensions, $\{\gamma^5,\bar\gamma^\mu\}=0$, and $[\gamma^5,\hat\gamma^\mu]=0$.
External momenta and the measured photon index are barred, while loop momenta and internal photon indices are $D$ dimensional.
{We use modified minimal subtraction ($\overline{\mathrm{MS}}$) for the UV poles and the finite evanescent subtractions below, without finite chiral-restoring counterterms.
The operators are defined with the canonical $\overline{\mathrm{MS}}$ photon field, while the final dipole coefficients are expressed in OS normalization.
The electron-dipole operator is}
\begin{equation}
O_{eD}=-\frac{i}{2}(\bar e\bar\sigma^{\mu\nu}\gamma^5e)\bar F_{\mu\nu},
\label{eq:electron-dipole-operator}
\end{equation}
and the $\tau$ dipole in Eq.~\eqref{eq:cp-odd-interactions} uses the same barred continuation.
{We define the evanescent operators by
\begin{equation}
\begin{aligned}
E_1&=(\bar\tau\hat\gamma^\nu i\gamma^5\tau)\partial^\mu F_{\mu\nu},\\
E_2&=-\frac{i}{2}(\bar\tau\hat\sigma^{\mu\nu}\gamma^5\tau)\hat F_{\mu\nu},\\
E_3&=\sum_{a,b,\rho}(\bar\tau\{i\gamma^5,\Gamma_{ab\rho}\}\tau)(\bar e\Gamma^{ab\rho}e).
\end{aligned}
\label{eq:evanescent-basis}
\end{equation}
Here $\Gamma_{ab\rho}=\gamma_{[a}\gamma_b\gamma_{\rho]}$, with antisymmetrization normalized by $1/3!$, and $\hat\sigma^{\mu\nu}=i[\hat\gamma^\mu,\hat\gamma^\nu]/2$.
We impose the finite subtraction~\cite{Dugan:1990df,Herrlich:1994kh}
\begin{equation}
[E_i]_R=[E_i]_{\overline{\mathrm{MS}}}-b_i[O_{eD}]_{\overline{\mathrm{MS}}}, \qquad \langle[E_i]_R\rangle_{d_e}=0.
\label{eq:evanescent-subtraction-condition}
\end{equation}
This condition is imposed at $\mu=m_\tau$ on the renormalized leading-power static projection, normalized by $\langle O_{eD}\rangle^{(0)}_{d_e}=1$.

\begin{figure}[tbp]
\color{black}
\centering
\includegraphics[width=\linewidth]{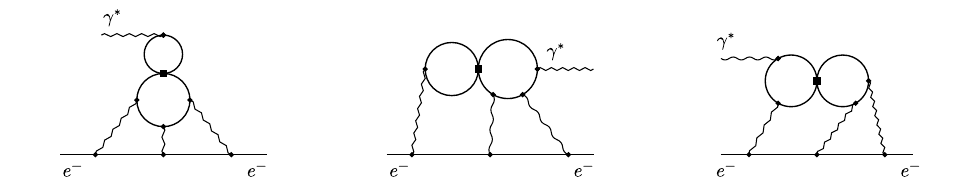}\\[2pt]
\makebox[\linewidth][l]{\hspace*{0.170\linewidth}\makebox[0pt]{(a)}\hspace*{0.336\linewidth}\makebox[0pt]{(b)}\hspace*{0.335\linewidth}\makebox[0pt]{(c)}}
\caption{Four-loop sources.
Panels (a) and (b): external- and internal-photon $1+3$ partitions; panel (c): $2+2$ partition between $\tau$ subloops.
Permutations are implied.}
\label{fig:fourloop-source}
\end{figure}

For three charged leptons, the finite-subtraction coefficients are
\begin{equation}
\begin{aligned}
b_1=m_e\left(\frac{\alpha}{4\pi}\right)^3\left(\frac{23}{9}+48\zeta(3)\right),\qquad
b_2=0,\qquad b_3=\frac{3\mu^{-\epsilon}m_e\mathrm{e}}{8\pi^2}\frac{\alpha}{\pi}.
\end{aligned}
\label{eq:finite-subtraction-coefficients}
\end{equation}
The subtraction condition holds through three loops for $E_{1,2}$ and two loops for $E_3$.
The $\overline{\mathrm{MS}}$-to-OS charge conversion affects only higher orders.
Appendix~\ref{app:evanescent} summarizes the source counterterms and the finite conversion.
}

Using Eq.~\eqref{leading-evolution} to express the $\tau$ coefficients at $\mu=m_\tau$, the $\tau$-induced electron EDM at leading order in $m_e$ is
\begin{equation}
\delta d_e=\frac{m_e}{m_\tau}\left(\frac{\alpha}{\pi}\right)^3 \left(\frac{19}{4}\zeta(3)-\frac{71}{24}\right) d_\tau(m_\tau) +\frac{\mathrm{e}m_eC_{SP}(m_\tau)}{8\pi^2}\left(\frac{\alpha}{\pi}\right)^3B_{SP}.
\label{electron-leading}
\end{equation}
{The physical EDM is $d_e^{\mathrm{phys}}=d_e(m_\tau)+\delta d_e$.
The quoted $B_{SP}$ uses the finite-subtraction prescription in Eq.~\eqref{eq:evanescent-subtraction-condition}.}
The explicit $\mu$ dependence in Eqs.~\eqref{tau-leading} and \eqref{electron-leading} has been deliberately suppressed by choosing $\mu=m_\tau$.
In the calculation, we have also retained $\mu\ne m_\tau$ and checked that both expressions agree with the results at $\mu=m_\tau$ at the retained order.
We retain electron-mass and external-momentum terms and all source classes in Fig.~\ref{fig:fourloop-source}.

Including both photon partitions, dimension-dependent region terms, and the evanescent conversion, we find
\begin{equation}
B_{SP}=I_{13}+I_{22}-\frac{19775}{41472}+\frac{\pi^4}{960}-\frac{11}{96}\zeta(3) =6.104.
\label{finite-matching-coefficient}
\end{equation}
Here $I_{13}$ and $I_{22}$ are the finite, dimensionless integrals of the internal-photon $1+3$ and $2+2$ sectors, before the displayed dimensional-region and evanescent corrections.
The external-photon class in Fig.~\ref{fig:fourloop-source}(a) gives no finite term at $\mu=m_\tau$: at leading power in $m_e/m_\tau$ and to linear order in the external photon momentum, its $\tau$ bubble attached to the probe photon reduces to a local dipole insertion, so after $\overline{\mathrm{MS}}$ subtraction it only yields $\log(\mu^2/m_\tau^2)$ times the part of the three-loop transfer in which the dipole couples to the external photon.

\begin{figure}[tbp]
\centering
\includegraphics[width=0.70\linewidth]{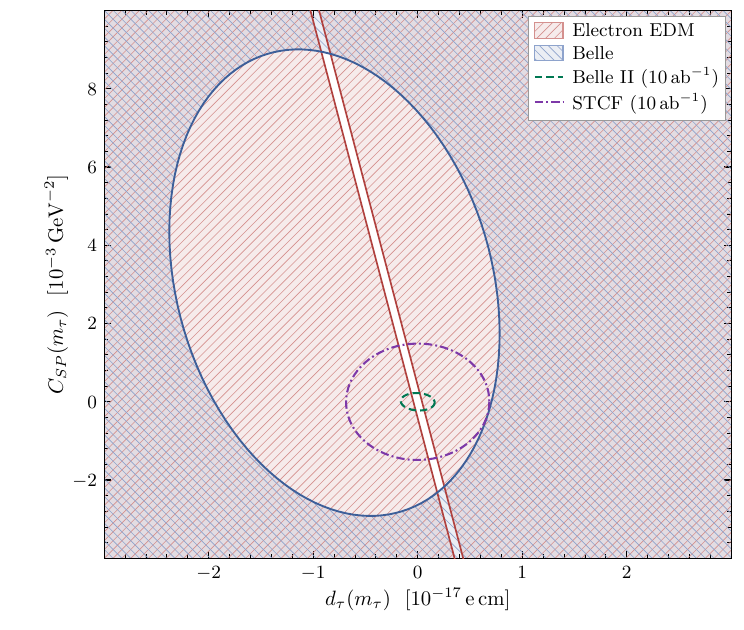}
\caption{Red hatching marks exclusion by the HfF$^+$ electron-EDM limit at 90\% C.L.~\cite{Roussy:2022cmp}, and blue hatching by the approximate Belle joint 95\% C.L. region for $\operatorname{Re}d_\tau$ and $\operatorname{Im}d_\tau$ at $\sqrt{s}=10.58\,\mathrm{GeV}$~\cite{Belle:2021ybo}; the white region is allowed by both.
Green dashed and purple dash-dotted contours show the Belle II ($\sqrt{s}=10.58\,\mathrm{GeV}$) and STCF ($\sqrt{s}=6.3\,\mathrm{GeV}$) projections of Refs.~\cite{He:2025ewk,Huang:2025ghw}.
All $\tau$ contours assume uncorrelated Gaussian uncertainties in $\operatorname{Re}d_\tau$ and $\operatorname{Im}d_\tau$, with $\Delta\chi^2=5.99$ for two parameters.}
\label{fig:joint-constraints}
\end{figure}

With $d_e=0$ at the matching boundary, Fig.~\ref{fig:joint-constraints} compares the electron and $\tau$ constraints in the plane of the two real source coefficients at $\mu=m_\tau$.
Assuming no independent electron--nucleon CP-odd interaction, the HfF$^+$ bound $|\delta d_e|<4.1\times10^{-30}\,\mathrm{e}\,\mathrm{cm}$ at 90\% confidence~\cite{Roussy:2022cmp} selects a narrow band around $\delta d_e=0$, but does not bound the two source coefficients independently when their contributions are allowed to cancel.
Belle~\cite{Belle:2021ybo} probes a complementary combination through the timelike form factor.
In particular, its absorptive part is proportional to $C_{SP}$ and cannot be canceled by a real local dipole at the leading order, so the $\tau$ measurement bounds the otherwise open cancellation direction.

\section{Conclusion}

We have computed the leading timelike $\tau$ electric dipole form factor and the four-loop electron-EDM matching induced by a CP-odd scalar--pseudoscalar four-$\tau$ interaction.
Under the assumed radiative counting, the four-$\tau$ source and the local $\tau$ dipole contribute at the same source order, and both the form factor and the electron matching are independent of the subtraction scale at this order.
With $d_e=0$ at the matching boundary, exact electron-EDM cancellation fixes the relative $\tau$-source contributions but leaves a timelike signal proportional to $C_{SP}$, with its momentum dependence and absorptive part determined by the $\tau$-pair cut.
The experimental electron bound broadens this cancellation line into a band, while $\tau$-pair measurements constrain the surviving direction through the real and imaginary parts of the timelike form factor.

\section*{Computational Methods}
The Dirac algebra and projections were performed in Mathematica with FeynCalc~\cite{Shtabovenko:2023idz,Shtabovenko:2025lxq}.
LiteRed2~\cite{Lee:2012cn,Lee:2013mka} was used for the integration-by-parts (IBP) reduction of the required evanescent subintegrals, and the finite remainder was evaluated numerically from convergent parameter integrals with our own Julia codes.
The source code for reproducing the results is publicly available in the Tau2EMatch repository.\footnote{\url{https://github.com/Ziyzou02/Tau2EMatch}}
The calculations, together with an independent cross-check using FIRE6~\cite{Smirnov:2019qkx} for IBP reduction, pySecDec~\cite{Borowka:2017idc,Heinrich:2023til} with FORM~\cite{Kuipers:2012rf} for a four-loop master integral, and Python with mpmath for high-precision numerical integration, were carried out by the AI models ChatGPT~6 Astra (OpenAI) and Claude Opus~5.5 (Anthropic) and were double-checked by the authors.

\acknowledgments
We thank Yong Du and ZhongLv Huang for helpful discussions.
This work was supported in part by the National Natural Science Foundation of China under Grant No. 12575096.

\FloatBarrier

\appendix
{
\section{Evanescent operators and finite subtractions}
\label{app:evanescent}

Let $O_{SP}=(\bar\tau\tau)(\bar\tau i\gamma^5\tau)$.
Here $n$ denotes the number of loops generating an evanescent source counterterm, $m$ denotes the number of loops with that counterterm inserted, and $(0)$ denotes tree level; coupling factors are included.
After subtracting the proper UV subdivergences,
\begin{equation}
\begin{aligned}
\left.\langle O_{SP}\rangle^{(n)}\right|_{\mathrm{UV,sub}} &=\sum_i\sum_{k=1}^n\frac{a_{i,k}^{(n)}}{\epsilon^k}\langle E_i\rangle^{(0)}+\cdots,\\
\left.\delta C_{E_i}^{(n)}\right|_{C_{SP}} &=-C_{SP}\sum_{k=1}^n\frac{a_{i,k}^{(n)}}{\epsilon^k}.
\end{aligned}
\label{eq:source-counterterms}
\end{equation}
The four-loop contributions satisfy $n+m=4$.
At the orders considered, the required residues are
\begin{center}
\renewcommand{\arraystretch}{1.8}
\setlength{\tabcolsep}{4pt}
\begin{tabular}{c|cc|cc}
 & $n$ & $m$ & $a_{i,1}^{(n)}$ & $a_{i,2}^{(n)}$\\ \hline
$E_1$ & 1 & 3 & $-\dfrac{\mathrm{e}}{24\pi^2}$ & $0$\\
$E_2$ & 1 & 3 & $-\dfrac{m_\tau\mathrm{e}}{8\pi^2}$ & $0$\\
$E_3$ & 2 & 2 & $-\dfrac{7}{27}\left(\dfrac{\alpha}{4\pi}\right)^2$ & $\dfrac{1}{18}\left(\dfrac{\alpha}{4\pi}\right)^2$
\end{tabular}
\end{center}
The $E_3$ entries are defined by matching the terms with one hatted index in Eq.~\eqref{eq:evanescent-basis}.
Representative counterterm insertions are shown in Fig.~\ref{fig:counterterm-insertions}.

\begin{figure}[!htbp]
\color{black}
\centering
\includegraphics[width=\linewidth]{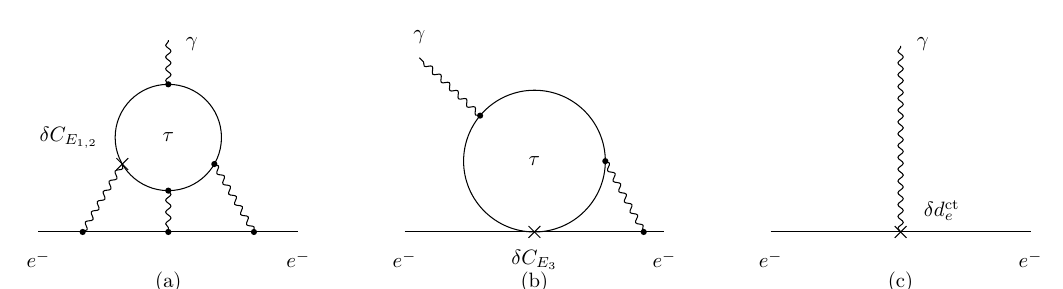}
\caption{Representative counterterm insertions: (a) $\delta C_{E_1}E_1$ or $\delta C_{E_2}E_2$ in a three-loop graph; (b) $\delta C_{E_3}E_3$ in a two-loop graph; (c) the local electron-EDM counterterm $\delta d_e^{\mathrm{ct}} O_{eD}$.
Crosses denote counterterm vertices; solid and wavy lines denote charged leptons and photons.
Permutations are implied.}
\label{fig:counterterm-insertions}
\end{figure}

\begin{figure}[tbp]
\color{black}
\centering
\includegraphics[width=\linewidth]{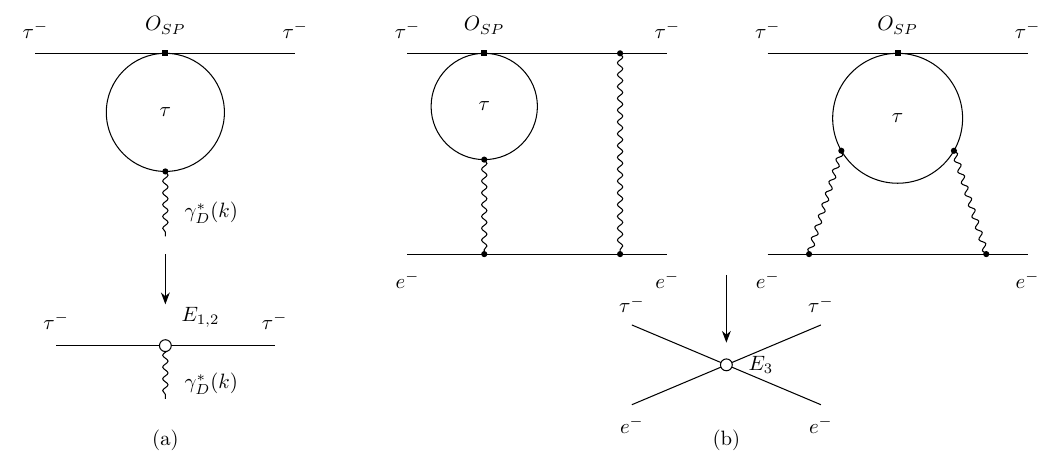}
\caption{Generation of evanescent operators by $O_{SP}$: (a) the one-loop source for $E_{1,2}$; (b) two-loop sources for $E_3$ from the $1+3$ (left) and $2+2$ (right) classes.
Exchange contractions are shown.
Squares denote $O_{SP}$, and open circles denote local operators.
Arrows indicate the UV poles after subdivergence subtraction.
In (a), $\gamma_D^*(k)$ is an off-shell, $D$-dimensional photon.}
\label{fig:evanescent-generation}
\end{figure}

The source diagrams that generate $E_{1,2,3}$ are shown in Fig.~\ref{fig:evanescent-generation}.
After UV subtraction, an evanescent insertion can leave a finite physical contribution,
\begin{equation}
\left.\langle[E_i]_{\overline{\mathrm{MS}}}\rangle^{(m)}\right|_{\mathrm{dipole}} =b_i^{(m)}\langle O_{eD}\rangle^{(0)}+\mathcal{O}(\epsilon).
\label{eq:evanescent-finite-projection}
\end{equation}
The finite subtraction in Eq.~\eqref{eq:evanescent-subtraction-condition}, with the coefficients given in Eq.~\eqref{eq:finite-subtraction-coefficients}, removes this contribution.
The accompanying physical counterterms preserve the reference $O_{SP}$ insertion and $B_{SP}$.
In the absence of independent renormalized evanescent sources at the matching scale, they also preserve the boundary condition $d_e(m_\tau)=0$.

\FloatBarrier
}

\bibliographystyle{tex/JHEP}
\bibliography{ref}

\end{document}